\documentclass[prb,twocolumn,showpacs,10pt,superscriptaddress]{revtex4}
\usepackage{graphicx}
\usepackage{dcolumn}
\usepackage{amsmath}
\usepackage{amsfonts}

\newcommand{\beq}{\begin{equation}}
\newcommand{\eeq}{\end{equation}}
\newcommand{\bqa}{\begin{eqnarray}}
\newcommand{\eqa}{\end{eqnarray}}
\newcommand{\nn}{\nonumber}

\newcommand{\erf}[1]{Eq.~(\ref{#1})}

\newcommand{\dg}{^\dagger}

\newcommand{\smallfrac}[2]{\mbox{$\frac{#1}{#2}$}}
\newcommand{\half}{\smallfrac{1}{2}}

\newcommand{\ito}{It\^o\ }
\newcommand{\str}{Stratonovich\ }

\newcommand{\sq}[1]{\left[ {#1} \right]}
\newcommand{\cu}[1]{\left\{{#1} \right\}}
\newcommand{\ro}[1]{\left( {#1}\right)}
\newcommand{\an}[1]{\left\langle{#1}\right\rangle}

\newcommand{\tr}[1]{\mathrm{Tr}\sq{ {#1}}}

\newcommand{\du}{\partial}
    \newcommand{\I }{{\cal I}}
    \newcommand{\W }{{\cal W}}
    \newcommand{\e }{\textrm{e}}
    
    \newcommand{\D}[1]{{\cal D}\sq{{#1}}}
    \newcommand{\J}[1]{{\cal J}\sq{{#1}}}
    \newcommand{\A}[1]{{\cal A}\sq{{#1}}}
    
    \newcommand{\h}[1]{{\cal H}\sq{{#1}}}
    \newcommand{\Ref}[1]{Ref.\ \onlinecite{#1}}
    \newcommand{\Fig}[1]{Figure\ \ref{#1}}
    \newcommand{\Sec}[1]{Sec.\ \ref{#1}}
    \newcommand{\sig}{\hat{\sigma}}
\begin{document}

\preprint{cond-mat/0401204}

\title{Quantum trajectories for the realistic measurement of a
  solid-state charge qubit}

\author{Neil P. Oxtoby}
\email{N.Oxtoby@griffith.edu.au}
\affiliation{Centre for Quantum Computer Technology, Centre for
  Quantum Dynamics, School of Science, Grif\mbox{}fith
  University, Nathan QLD 4111, Australia}
\affiliation{Centre for Quantum Computer Technology, School of
  Physical Sciences, The University of Queensland, St. Lucia QLD
  4072, Australia}

\author{P. Warszawski}
\affiliation{Centre for Quantum Computer Technology, Centre for
  Quantum Dynamics, School of Science, Grif\mbox{}fith
  University, Nathan QLD 4111, Australia}

\author{H. M. Wiseman}
\email{H.Wiseman@griffith.edu.au}
\affiliation{Centre for Quantum Computer Technology, Centre for
  Quantum Dynamics, School of Science, Grif\mbox{}fith
  University, Nathan QLD 4111, Australia}

\author{He-Bi Sun}
\affiliation{Centre for Quantum Computer Technology, School of
  Physical Sciences, The University of Queensland, St. Lucia QLD
  4072, Australia}
\affiliation{Centre for Quantum Computer Technology, Centre for
  Quantum Dynamics, School of Science, Grif\mbox{}fith
  University, Nathan QLD 4111, Australia}

\author{R. E. S. Polkinghorne}
\affiliation{Centre for Quantum Computer Technology, School of
  Physical Sciences, The University of Queensland, St. Lucia QLD
  4072, Australia}

\begin{abstract}
We present a new model for the continuous measurement of a coupled
quantum dot charge qubit.  We model the \textit{effects} of a
realistic measurement, namely adding noise to, and filtering, the
current through the detector.  This is achieved by embedding the
detector in an equivalent circuit for measurement.  Our aim is to
describe the evolution of the qubit state \textit{conditioned} on the
macroscopic output of the external circuit.  We achieve this by
generalizing a recently developed quantum trajectory theory for
realistic photodetectors
[P. Warszawski, H. M. Wiseman and H. Mabuchi, Phys. Rev. A \textbf{65}
023802 (2002)] to treat solid-state detectors.  This yields stochastic
equations whose (numerical) solutions are the ``realistic quantum
trajectories'' of the conditioned qubit state.  We derive our general
theory in the context of a low transparency quantum point contact.
Areas of application for our theory and its relation to previous work
are discussed.
\end{abstract}
\date{December 23, 2004}

\pacs{73.23.Hk, 03.67.Lx}
\keywords{Quantum trajectory, conditional dynamics, open quantum
  system, realistic detection, quantum computation, non-Markovian}

\maketitle

%%%%%%%%%%%%%%%%%%%%%%%%%%%%%%%%%%%%%%%%%%%%%%%%%%%%%%%%%%%%%%%%%%%%%%%%%
\section{Introduction}
\label{sec:intro}
The field of research that surrounds the quest for a
large-scale quantum computer is very exciting.  At present,
solid-state proposals~\cite{kane,dots,privman,imamoglu,vrijen} seem
promising.  The ability to read out the state of the quantum bits
(qubits) of information is of obvious importance in any quantum
computational scheme.  In this paper we consider continuous
measurement of the state of a pair of coupled quantum dots (CQDs)
occupied by a single excess electron.  This constitutes a
\textit{charge} qubit.  It is worth mentioning that spin
qubits\cite{kane,dots,privman,imamoglu,vrijen} are considered more
favorably for solid-state quantum computation %in the long term 
due to their relatively long coherence times,\cite{longcoherence} but
read-out may have to be performed via charge qubits using
spin-to-charge conversion.\cite{kane,burkloss2002,delftspin}

The evolution of solid-state qubits subject to continuous measurement
has received considerable theoretical consideration
recently.\cite{gurvitz, Shnir98, korotkov99, korotkov01b, setpaper,
  qpc1, korotkov01c, goanmilburn, korotkov03, korotkov03review,
  gurvitzPRL}
 Single realizations of the continuous measurement of a solid-state
CQD qubit, known as \textit{conditional} (or selective) evolution,
have been treated by a number of
groups.\cite{korotkov99, korotkov01b, setpaper, qpc1, korotkov01c,
  goanmilburn, korotkov03, korotkov03review, gurvitzPRL}
%also by refs 13-18 in Korotkov's
%  "Noisy QM of SS Qubits: Bayesian approach"
These works conditioned the qubit evolution on quantum processes (such
as tunneling) at the scale of a mesoscopic detector.  They did not
consider conditioning on the macroscopic current that is realistically
available to an observer.  In particular, they ignored the noisy
filtering characteristic of the external circuit, including an
amplifier.  It is worth noting that non-idealities have been
considered in some of these works.  \Ref{setpaper} considered a
detector with excessive back-action.  \Ref{korotkov03} did this also,
and also considered extra classical noise, phenomenologically.
\Ref{goanmilburn} considered ``inefficient'' measurements.  None of
these considered filtering.

In this paper we consider the evolution of a solid-state qubit
conditioned on the output available to a realistic observer, which has
been filtered and degraded (i.e. made more noisy) by an external
circuit. That is, we are interested in the evolution of the system
conditioned on information available to an observer, not on the
microscopic events occurring within the detector to which a real
observer has no direct access. Being able to determine the state of a
quantum system conditioned on actual measurement results is essential
for understanding and designing feedback
control.\cite{WMilPRL93, WisePRL95, DtyJac99, DohertyPRA00,
  ArmenPRL02, opencontrol, SmithPRL02, korotkov01b, RuskovFB02}
As well as being intrinsically interesting, this is also expected to
be important in quantum computing, both for state preparation and
quantum error correction.\cite{AhnWisMil03,SarAhnJacMil04,AhnWisJac04}

A \textit{quantum trajectory}\cite{opensystems, WiseMilQT93, WiseQTQM,
setpaper} describes the Markovian stochastic evolution of an open
quantum system conditioned on continuous monitoring of its output by
a \textit{bare} detector.  A ``bare'' detector is one which does not
include the noisy filtering characteristic of realistic measurements.
In an experiment the output from this detector is filtered through
various noisy electronic devices.  Due to the finite bandwidth of all
electronic devices, the evolution of the conditional state of the
quantum system must be non-Markovian.  A general method of describing
this evolution was presented in recent 
papers\cite{warwismab, photodetection1} by two of us in the context of
photodetection, where it was applied to an avalanche photo-diode and a 
photo-receiver.  In the present paper the theory of
\Ref{photodetection1} is applied to a solid-state detector -- the low
transparency quantum point contact\cite{Field,gurvitz} (QPC), or
tunnel junction, which is an ideal detector.\cite{korotkov99}  
In our approach an equivalent circuit is used to model the effects of
a realistic measurement.  Note that for clarity we will use the
terminology \textit{detector} for a bare detector and
\textit{measurement device} for a detector embedded within a
measurement circuit.

The paper is organized as follows.  We begin in the next section by
describing our models for the qubit and the QPC (including the
monitored qubit's conditional and average dynamics in the bare
detector case).  We then introduce and analyze our equivalent circuit
for realistic measurement in \Sec{sec:circuit}.  The method of
deriving realistic quantum trajectories is presented in
\Sec{sec:derivation}, in the context of a QPC.  We discuss our results
in \Sec{sec:discussion} and conclude in \Sec{sec:conclusion} with a
summary, comparison with previous work, and prospects for future
work.

%%%%%%%%%%%%%%%%%%%%%%%%%%%%%%%%%%%%%%%%%%%%%%%%%%%%%%%%%%%%%%%%%%%%%%%
\section{System}
\label{sec:system}
In this section we describe the models for the qubit and the detector.
Using a master equation formalism we present the conditional and
ensemble average dynamics of the qubit state when measured by a low
transparency QPC.  The conditional qubit dynamics in the bare
measurement case are represented by a \textit{stochastic} master
equation.  We choose to present stochastic differential equations in
the \ito formalism rather than the alternative \str
formalism.\cite{CWGhbook}  Extension of the theory to our more
realistic measurement case occurs later.

\begin{figure}[ht]
\begin{center} \includegraphics[width=0.3\textwidth]{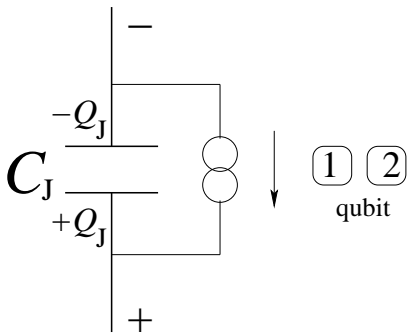}
\end{center}
\caption{An equivalent circuit for a low transparency QPC or
tunnel junction and nearby charge qubit. The arrow indicates the
direction of electron tunneling through the QPC (represented by a
current source).
\label{fig:qpc} }
\end{figure}

\Fig{fig:qpc} is a schematic representation of the CQD qubit and
nearby low transparency QPC or tunnel junction.  
%The tunnel junction,
%modeled with a small capacitor $C_{\rm J}$, separates source and drain
%electron reservoirs.  Electron tunnel events through the junction are
%respresented by a current source.  
The CQDs (labeled 1 and 2) are occupied by a single excess electron,
the location of which determines the logical state of the qubit.  We
assume that each quantum dot has only one single-electron energy
level available for occupation by the qubit electron.  These energy
levels are denoted by $E_1$ and $E_2$.  

%%%%%%%%%%%%%%%%%%%%%%%%%%%%%%%%%%%
%%%%%%%%%%%%%%%%%%%%%%%%%%%%%%%%%%%

Using the convention of $\hbar =1$ (as we will for the entire paper),
the total Hamiltonian for the qubit can be written as
\begin{equation}\label{eq:Hqb}
\hat{H}_{qb} = E_1\hat{c}_1\dg\hat{c}^{\phantom{\dagger}}_1
             + E_2\hat{c}_2\dg\hat{c}^{\phantom{\dagger}}_2
 + \frac{\Omega_0}{2}\ro{\hat{c}\dg_1\hat{c}^{\phantom{\dagger}}_2
 + \hat{c}\dg_2\hat{c}^{\phantom{\dagger}}_1} \ ,
\end{equation}
where $\Omega_0$ is the co-efficient of tunneling between the qubit
dots % (the Rabi frequency when $E_1=E_2$) 
and
$\hat{c}^{\phantom{\dagger}}_1$ ($\hat{c}^{\phantom{\dagger}}_2$) is
the Fermi annihilation operator for the single available electron
state within the qubit dot labeled 1 (2).  The qubit electron tunnels
between the two dots at the Rabi frequency $\Omega = \sqrt{\Omega_0^2
  + \varepsilon^2}$, where $\varepsilon \equiv E_1-E_2$ is the
asymmetry in the CQD energy levels.

The state of a measured quantum system is affected by the detector in
two ways.  First, there is the measurement back-action caused by
their mutual interaction.  Second, if the output of the detector is
observed, then the state of the system is conditioned by the
stochastic outcomes.  We describe the conditional dynamics of the
measured qubit, including the measurement back-action, using a
stochastic approach.  In the case of measurement with a bare ideal
detector, the state of the qubit is conditioned by electron tunneling
events through the detector which constitute an idealized output
current.  For such an ideal detector the measurement back-action is
quantum-limited, also called Heisenberg-limited.\cite{qmBragKhal}

A number of formalisms exist that describe the evolution of a measured
quantum system conditioned on a particular measurement result from the
detector.  The conditional dynamics of continuously measured CQD
systems have been treated by Bloch-type
equations,\cite{gurvitz,gurvitzPRL} quantum trajectory
theory\cite{setpaper,qpc1,goanmilburn} and a Bayesian
formalism.\cite{korotkov99, korotkov01b, korotkov01c, korotkov03,
  korotkov03review}
This Bayesian formalism has been shown to coincide with the
quantum trajectory formalism with only notational differences (see the
appendix of \Ref{qpc1}).  All three formalisms coincide for the
ensemble average dynamics of the measured CQD system.  In the
stochastic approach, the (Markovian) conditional dynamics of the
measured qubit state is described with a stochastic master equation.
This generates a ``quantum trajectory'', so called because it tracks
the state of the quantum system in time.  We also present the ensemble
average master equation.

The equivalent circuit for the QPC coupled to the qubit is shown in
\Fig{fig:qpc}.  We represent the tunnel junction by a
capacitance $C_{\rm J}$, which contains the charge $Q_{\rm J}$.
The stochastic electron tunneling events through the junction are
represented by a current source.  The location of the CQD electron
changes the height of the potential barrier in the QPC and
consequently the current through it, thus providing the means to
measure the qubit state.  For simplicity, we assume that electrons
tunnel only from source to drain.  This tunneling occurs at two
different rates, namely $r$ and $r'$, which correspond to the near
(dot 1) and far (dot 2) CQD being occupied, respectively.

% These electrons could in fact be affected by the qubit electron's
% location -- perhaps a phase shift, or something else -- but we
% assume the only effect of the qubit electron on the circuit is to
% raise and lower the QPC barrier potential.
The ensemble average master equation for the qubit state, $\rho$, when
measured by a low transparency QPC, or similar single tunnel-junction
device, is\cite{gurvitz,korotkov99,setpaper,qpc1}
\begin{equation}
\label{eq:MEqpc}
\frac{d\rho}{dt} = -i[\hat{H}_{qb},\rho(t)]
+ \D{{\cal T} + {\cal X}\hat{n}}\rho(t)
\equiv {\cal L} \rho\ .
\end{equation}
Here $\hat{n} = \hat{c}\dg_1\hat{c}^{\phantom{\dagger}}_1$ is the
occupation of the near dot.  The Lindblad superoperator ${\cal D}$
represents the irreversible part of the qubit evolution -- the
decoherence.  It is defined in terms of two other superoperators,
${\cal J}$ and ${\cal A}$:
\begin{equation}\label{eq:D}
\D{\hat{X}}\rho \equiv \J{\hat{X}}\rho - \A{\hat{X}}\rho\ ,
\end{equation}
where ${\cal J}$ (the `jump' superoperator) and ${\cal A}$ (the
anti-commutating superoperator) are defined by
\begin{eqnarray}
\label{eq:J}
\J{\hat{X}}\rho &\equiv& \hat{X} \rho\hat{X}\dg \ ,\\
\label{eq:A}
\A{\hat{X}}\rho &\equiv& \half \ro{\hat{X}\dg \hat{X}\rho +
  \rho\hat{X}\dg \hat{X}}\ .
\end{eqnarray}
These superoperators, introduced in \Ref{WiseMilQT93}, are used
commonly in quantum optics measurement theory.

For simplicity we assume real tunneling amplitudes whereby
\begin{equation}
{\cal T}^2 = r'\ , \quad \ro{\cal T + X}^2 = r \ ,
\label{eq:tunamps_qpc}
\end{equation}
which implies that ${\cal X}<0$.  Complex tunneling amplitudes are
allowed in the model of \Ref{qpc1} and the generalization here would
be straightforward.

A realistic observer may not be able to tell when a tunneling event
through the QPC occurs.  However, we argue that in principle this
information would be contained in the movement of the Fermi sea
electrons in the leads attached to the QPC.  Thus, we can legitimately
represent the \textit{conditional} evolution of $\rho(t)$ (denoted
with a superscript $\mu$) in terms of these microscopic events. 
Using the method for quantum jumps introduced in \Ref{setpaper}, this
conditional evolution is described by the following \ito stochastic
master equation\cite{qpc1}
\begin{eqnarray}
d\rho^{\mu}(t) &=& dN(t)\cu{\frac{\J{{\cal T} + {\cal
	X}\hat{n}}}{
%\tr{\J{{\cal T} + {\cal X}\hat{n}}\rho^{\mu}(t)
\mathrm{E}[dN(t)]/dt
}
 - 1}\rho^{\mu}(t) 
\nn \\
&&- dt~ \h{ \half \hat{R} +i\hat{H}_{qb}}\rho^{\mu}(t)\ ,
%&&+ dt\bigg\{-\A{{\cal T}+{\cal X}\hat{n}}\rho^{\mu}(t) 
%+ \frac{\mathrm{E}[dN(t)]}{dt}\rho^{\mu}(t) \bigg. \nn \\
%&&\bigg. \qquad -i[\hat{H}_{qb},\rho^{\mu}(t)]\bigg\}\ , 
\label{eq:SMEqpc}
\end{eqnarray}
where we have introduced the classical point process $dN(t)$ that
represents the number (either 0 or 1) of electron tunneling events
through the QPC in an infinitesimal time interval $[t,t+dt)$.  The
expectation (ensemble average) value of $dN(t)$ is
\begin{equation}
\mathrm{E}[dN(t)] =
dt~ \tr{\J{{\cal T} + {\cal X}\hat{n}}\rho^{\mu}(t)} \ .
\label{eq:E[dN]qpc}
\end{equation}
In \erf{eq:SMEqpc} we have also introduced the convenient
superoperator ${\cal H}$ and operator $\hat{R}$ which are defined by
\begin{eqnarray}
\label{eq:Hcal}
\h{\hat{X}}\rho &\equiv& \hat{X}\rho + \rho\hat{X}\dg -
  \tr{\ro{\hat{X}+\hat{X}\dg }\rho}\rho\ ,\\
\label{eq:Rqpc}
\hat{R} &\equiv& \ro{{\cal T}+{\cal X}\hat{n}}^2 \ .
\end{eqnarray}

It can be seen from the master equation (\ref{eq:MEqpc}) that the
minimum tunneling rate through the QPC, $r$, occurs when qubit dot 1
is occupied ($n=1$).  This is due to maximum electrostatic repulsion
between the qubit electron and electrons in the QPC vicinity.
Accordingly, the maximum QPC tunneling rate $r'$ occurs when $n=0$.
These tunneling rates could be functions of the voltage across the
detector, which we consider as changing with time.  However, this
would necessarily mean that the measured qubit's evolution cannot be
described by the quantum master equation formalism.  To allow for this
would be to go beyond what anyone has done in this area.

%Similar master equations to (\ref{eq:MEqpc}) and (\ref{eq:SMEqpc}) are
%presented in \Ref{setpaper} for measurement of a CQD qubit by a highly
%asymmetric single electron transistor.  This device could easily be
%incorporated into our theory.

%%%%%%%%%%%%%%%%%%%%%%%%%%%%%%%%%%%%%%%%%%%%%%%%%%%%%%%%%%%%%%%%%
\section{Equivalent circuit for realistic measurement}
\label{sec:circuit}
%%%%%%%%%%%%%%%%%%%%%%%%%%%%%%%%%%%%%%%%%%%%%%%%%%%%%%%%%%%%%%%%%

\begin{figure}[ht]
\begin{center} \includegraphics[width=0.4\textwidth]{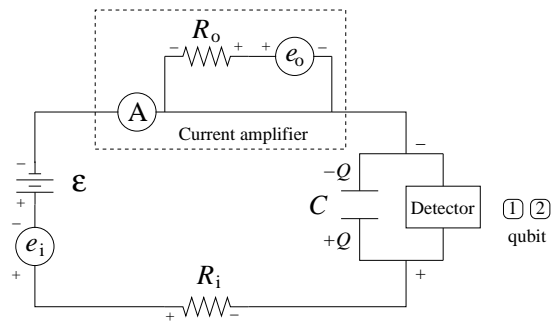}
\end{center}
\caption{A schematic of our equivalent circuit for the realistic
measurement of the state of a CQD charge qubit.
\label{fig:circuit}}
\end{figure}
%%%%%%%%%%%%%%%%%%%%%%%%%%%%%%%%%%%%%%%%%%%%

Our equivalent circuit for realistic measurement of the CQDs is shown
in \Fig{fig:circuit}.  We emphasize that this circuit models
\textit{effects} of realistic measurement (additional classical 
noise and filtering of the signal), \textit{not} an actual
experimental apparatus.

The circuit is biased by a non-ideal DC voltage consisting of a
noiseless voltage $\varepsilon$ and a noisy voltage source
$e_\mathrm{i}$.  This (white) noise source could be considered as the
Johnson-Nyquist noise from the equivalent circuit resistance
$R_\mathrm{i}$ at some effective temperature $T_{\rm i}$.  We
emphasize again that this is a model only and need not correspond to a
real temperature in an actual apparatus.  The small current through the
detector is amplified, then measured.  In this process an observer
will see white noise in addition to the current through the
detector. This is modeled by adding a noisy output current
$e_\mathrm{o}/R_\mathrm{o}$ to the signal from the detector prior to
measurement by a perfect ammeter, yielding the current $\I$.  
%Any filtering that may occur due to capacitance within the amplifier
%is ignored.  
The \textit{parasitic} capacitance $C$ across the detector is due to
the large cross-sectional area of the leads relative to the detector
junction.

Again, it is important to note that the circuit components
are \textit{not} necessarily representative of an actual experimental
setup.  For example, an amplifier does not consist of a noisy voltage
and a resistor, rather the observed \textit{effect} of amplification
of the current through the detector can be modeled as the addition of
an output noise $e_\mathrm{o}/R_\mathrm{o}$ to the current through the
detector.  Although our description of the circuit is rather simple,
we believe that it is a reasonable starting point that models some
essential effects of a realistic measurement.  Future improvements to
this circuit model could include considering an actual circuit from an
experiment.

We analyze the equivalent circuit with the low transparency QPC as
detector and produce expressions for the measured current $\I (t)$ and
the time evolution of the parasitic capacitor charge $Q(t)$.  The
variable $Q(t)$ is used to describe the state of the circuit part of
the measurement device.

For the moment, ignore tunneling through the QPC.  Analysis of the
measurement circuit using Kirchhoff's electrical circuit laws yields
the following \ito differential equation for the increment in $Q$ (the
charge on the parasitic capacitor) due to the circuit components
\begin{equation}
dQ(t) = \left( -\frac{Q(t)}{R_\mathrm{i}C}
+ \frac{\varepsilon}{R_\mathrm{i}}
+ \frac{e_\mathrm{i}}{R_\mathrm{i}} \right)dt\ . \label{eq:dQ1}
\end{equation}

Similar analysis yields an expression for the measured current as
a function of time:
\begin{equation}
\I (t) = - \frac{Q(t)}{R_\mathrm{i}C} 
         + \frac{\varepsilon}{R_\mathrm{i}} 
         + \frac{e_\mathrm{i}}{R_\mathrm{i}} 
         + \frac{e_\mathrm{o}}{R_\mathrm{o}}\ .
\label{eq:I1}
\end{equation}

For the purposes of our work it is useful to express the
(Johnson-Nyquist) noise sources $e_\mathrm{i}$ and $e_\mathrm{o}$ in
terms of stochastic increments.  In the steady state, Johnson-Nyquist
voltage noise is \textit{white} noise and has a flat spectrum
\begin{equation}
S = 2k_\mathrm{B}TR\ , \label{eq:Sflat}
\end{equation}
where $T$ is the temperature of the resistor $R$ and $k_\mathrm{B}$ is
Boltzmann's constant.  The current spectrum (`spectral density')
definition\cite{yalespie} used here is
\begin{equation}
S(\omega ) = \int_{-\infty}^{\infty}\exp\sq{i\omega \tau}G(\tau)d\tau\ ,
%= 2\int_0^{\infty}\cos (\omega \tau )G(\tau )d\tau\ ,
\label{eq:spectrum}
\end{equation}
where $G(\tau)$ is the two-time autocorrelation function of the
measured current.\footnote{The engineering definition of the spectral
  density is a factor of two larger than \erf{eq:spectrum} due to the
  addition of the negative and positive frequency components of the
  spectral density. %, which are equal when considering physical
		    %quantities. 
  See \Ref{yalespie} for a detailed consideration of this.}
Obviously a current flow is not an equilibrium situation, but for
reasonable bias voltages the approximation of \erf{eq:Sflat} remains
valid.\cite{HeBiGerard}  For simplicity, we take the flat spectra of
the input and output voltage noises to be
$2D_\mathrm{i}R^2_\mathrm{i}$ and $2D_\mathrm{o}R^2_\mathrm{o}$,
respectively.  This allows us to write Eqs.~(\ref{eq:dQ1}) and
(\ref{eq:I1}) in terms of the input and output Wiener processes, 
$dW_\mathrm{i}(t)$ and $dW_\mathrm{o}(t)$, as
\begin{eqnarray}
dQ(t) &=& \left( -\frac{Q(t)}{R_\mathrm{i}C} 
                 +\frac{\varepsilon}{R_\mathrm{i}} \right)dt
         +\sqrt{D_\mathrm{i}}dW_\mathrm{i}(t) \ ,
\label{eq:dQ}\\ 
\I (t) &=& -\frac{Q(t)}{R_\mathrm{i}C} 
+ \frac{\varepsilon}{R_\mathrm{i}}
+ \sqrt{D_\mathrm{i}}\frac{dW_\mathrm{i}(t)}{dt}
+ \sqrt{D_\mathrm{o}}\frac{dW_\mathrm{o}(t)}{dt}\ , \nn \\
\label{eq:I}
\end{eqnarray}
where $D_\mathrm{i}=2k_{\rm B}T_\mathrm{i}/R_\mathrm{i}$ and
$D_\mathrm{o} = 2k_{\rm B}T_\mathrm{o}/R_\mathrm{o}$. 
These expressions correct the expressions in
Refs.~\onlinecite{warwismab},~\onlinecite{photodetection1}
and~\onlinecite{OxJPCM03} from $4k_\mathrm{B}T/R$ to
$2k_\mathrm{B}T/R$.  The Wiener increment is related to Gaussian white
noise $\xi (t)$ by $dW(t) = \xi (t)dt$.\cite{CWGhbook}

Now consider a single electron tunneling event through the QPC
($dN(t)=1$).  The charge on the parasitic capacitor will change by an
amount $\e  dN(t)$, where $\e$ is the charge on an electron.  This
gives
\begin{equation}
dQ(t) = \sq{-\alpha Q(t) + \beta}dt
+ \sqrt{D_\mathrm{i}}dW_\mathrm{i}(t) + \e dN(t),
\label{eq:dQfinalqpc}
\end{equation}
where we have introduced the simplifying notations
$\alpha = 1/R_\mathrm{i}C$ and $\beta = \varepsilon /R_\mathrm{i}$.
The solution to this differential equation gives the value of $Q(t)$
that may be substituted into \erf{eq:I} to give a lengthy expression
for the measured current.\cite{OxJPCM03}

%%%%%%%%%%%%%%%%%%%%%%%%%%%%%%%%%%%%%%%%%%%%%%%%%%%%%%%%%%%%%%%%%%%%%
\section{Derivation of realistic quantum trajectories}
\label{sec:derivation}
The derivation of realistic quantum trajectories follows a number of
well defined steps as presented for photodetectors in
\Ref{photodetection1}.  We refer the reader to \Ref{photodetection1}
for specific details of the derivation steps and only present the
essential points and details that are unique to the solid-state
situation.  Note however that we use a somewhat simpler derivation,
using the Zakai equation in \Sec{subsec:ze} rather than the
Kushner-Stratonovich equation of \Ref{photodetection1}.

\subsection{Stochastic differential Chapman-Kolmogorov equation}
\label{subsec:sdcke}
\erf{eq:dQ} describes the evolution of the circuit state for
situations where $Q(t)$ is known.  A realistic observer will not have
direct access to the precise value of $Q(t)$ due to the randomness of
the microscopic events occurring within the device.  We therefore
require an equation for the evolution of the probability distribution
for $Q$, written $P(q)$.  Following the procedure outlined in
\Ref{photodetection1}, we obtain the stochastic differential
Chapman-Kolmogorov (SDCK) equation for the evolution of $P(q)$:
\begin{eqnarray} dP^{\mu}(q) &=& dt\ro{-\frac{\du }{\du q}m
  + \frac{D_\mathrm{i}}{2} \frac{\du ^{2}}{\du q^{2}} 
- \sqrt{D_\mathrm{i}}\frac{\du}{\du q}dW_\mathrm{i}
}P(q) \nn \\
%&&- \sqrt{D_\mathrm{i}}\frac{\du}{\du q}dW_\mathrm{i}P(q) \nn \\
&&+ dN \sq{ P\ro{q-\e} - P(q)} \ ,
\label{eq:realqt:sdcke}
\end{eqnarray}
where $m=-\alpha q + \beta$.  This equation gives the increment in the
probability distribution for the charge on the parasitic capacitor
conditioned by the unobserved microscopic events ($\mu$) occurring
within the measurement device.

\subsection{Zakai equation}
\label{subsec:ze}
The state of the circuit part of the measurement device is now
represented by the probability distribution $P(q)$ that was introduced
in the previous section.  The state of this classical system changes
upon measurement and so $P(q)$ must be updated.  The best estimate of
the new probability distribution representing the conditioned state of
the measurement device, given a measurement result $\I $, is found
using Bayesian analysis\cite{bayes} to be
\begin{equation}
\tilde{P}(q|\I ) = \frac{P(\I |q)P(q)}{\Lambda(\I )}~, \label{eq:bayes1}
\end{equation}
where \mbox{$\Lambda(\I )= P(\I |q=\beta/\alpha)$}.  
Here $\tilde{P}(q|\I )$ is read `the probability of $q$ given $\I $'.
The tilde denotes an unnormalized distribution and the value of
$q=\beta/\alpha$ is chosen for convenience.  The Zakai equation tells
us how to update the probability distribution $P(q)$ when the
measurement result $\I $ is obtained.  The quantity $P(\I |q)$ is the
probability of obtaining the result $\I$ given that the state is $q$.
We will use the simpler notation $P_q(\I)\equiv P(\I|q)$, where the
subscript denotes the result upon which the conditioning is
performed.  $\Lambda(\I)$ can be thought of as the \textit{ostensible}
probability distribution,\cite{WiseQTQM} as opposed to the actual
probability distribution
\begin{equation}
P(\I ) = \int dq P(\I |q)P(q) = \Lambda(\I)\int dq \tilde{P}(q|\I)\ ,
\label{eq:P=Lambda}
\end{equation}
which replaces $\Lambda(\I)$ in the expression for the normalized
distribution $P(q|\I)$.\cite{photodetection1}

From our expression for the measured current, \erf{eq:I}, $P_q(\I)$ is
a Gaussian distribution with a variance of $\nu = (D_\mathrm{i} +
D_\mathrm{o})/dt$ and a mean of $m=-\alpha q + \beta$.  
Thus, \erf{eq:bayes1} gives the the Zakai equation (to order $dt$):
\begin{equation}
\tilde{P}_\I(q) = \sq{ 1 + \I dt\frac{m}{D_\Sigma}} \tilde{P}(q)
\ ,\label{eq:ze}
\end{equation}
%%%%%%%%%%%%%%%%%%%%%%%
% NOTE: The tilde on the P on the RHS of this equation is only the
% general case.  For our work it is actually normalised.  When
% multiplied by the \I dt term, it is unnormalised.
%%%%%%%%%%%%%%%%%%%%%%%
where we have defined \mbox{$D_\Sigma = D_\mathrm{i} + D_\mathrm{o}$}
for convenience.  Note that $\I$ has the ostensible distribution
$\Lambda(\I )= \exp \sq{-\I^2/2\nu} / \sqrt{2\pi \nu}$.

\subsection{Combining the stochastic increments}
\label{subsec:combine}
Our description of the stochastic conditional evolution of the
measurement device is found by combining the increments
$dP^{\mu}(q)$ and $d\tilde{P}_{\I}(q)$ given in the previous two
subsections.  The stochasticity of these two increments is related as
the input noise $dW_\mathrm{i}$ plays a role in both.  For this reason
we must combine them into one increment using
\begin{equation}
\tilde{P}(q) + d\tilde{P}^{\mu}_{\I}(q)
= \sq{1 + \I dt \frac{m}{D_\Sigma}}\sq{P(q) + dP^{\mu}(q)}\ ,
\label{eq:combine}
\end{equation}
rather than by simply adding them together.  Remembering that we will
eventually average over unobserved processes, the input noise needs to
be separated into observed and unobserved parts.  We express this as
\begin{equation}\label{eq:inputnoise}
dW_\mathrm{i} = a \I dt  + b dW' + c dt\ ,
\end{equation}
where
\begin{equation}\label{eq:uncorrelated}
dW'\I dt = 0 \ .
\end{equation}
Here $a$, $b$ and $c$ are as yet undetermined expressions and $dW'$ is
unobserved, normalized white noise that is unrelated to the known
output $\I $.  When averages are taken, $dW'$ will be averaged over
and $\I $ kept.  The observed output [\erf{eq:I}] can be expressed as
\begin{equation}
\I dt = m dt %\ro{-\alpha Q + \beta}dt
       + \sqrt{D_\mathrm{i}}dW_\mathrm{i}
       + \sqrt{D_\mathrm{o}}dW_\mathrm{o}\ .
\label{eq:observedI}
\end{equation}
Using Eqs.~(\ref{eq:uncorrelated}) and (\ref{eq:observedI}) gives the
expression for $dW'$:
\begin{equation}\label{eq:dW'}
dW' = \frac{\sqrt{D_\mathrm{i}}dW_\mathrm{o}
          - \sqrt{D_\mathrm{o}}dW_\mathrm{i}}
           {\sqrt{D_\Sigma}}\ .
\end{equation}
Using this in \erf{eq:inputnoise} and equating the left and right
hand sides allows $a$, $b$ and $c$ to be determined.  Substitution of
$a$, $b$ and $c$ back into \erf{eq:inputnoise} yields
\begin{eqnarray}
dW_\mathrm{i} &=& \frac{\sqrt{D_\mathrm{i}}}{D_\Sigma}\I dt
- \sqrt{\frac{D_\mathrm{o}}{D_\Sigma}}dW'
- \frac{\sqrt{D_\mathrm{i}}}{D_\Sigma}m dt\ .
\label{eq:inputnoise2}
\end{eqnarray}

Using this result and the SDCK equation [\erf{eq:realqt:sdcke}] in
\erf{eq:combine} gives
\begin{eqnarray}
P(q)+d\tilde{P}^\mu_\I(q) &=&\left\{
1+dt\ro{ -\frac{\du }{\du q}m
     ~+~ \frac{D_\mathrm{i}}{2}\frac{\du^2}{\du q^2}}
\right.
\nn\\
&&\left.
~+~\ro{m- D_\mathrm{i}\frac{\du}{\du q}}\frac{\I dt}{D_\Sigma}
\right.\nn \\
&& \left.
~+~\sqrt{\frac{D_\mathrm{i}D_\mathrm{o}}{D_\Sigma}}
\frac{\du}{\du q} dW'
\right\} P(q)\nn \\
&& ~+~ dN\sq{P(q-\e) - P(q)}\ .
\label{eq:combined}
\end{eqnarray}
This result represents the evolution of the circuit state conditioned
on both the microscopic events occurring within the device and the
observed current $\I$.

\subsection{Joint stochastic equation}
\label{subsec:joint}
The stochastic state of the joint classical-quantum system is found by
forming the new conditional quantity
\begin{equation}
\tilde{\rho}^{\mu}_{\I}(q) = \tilde{P}^{\mu}_{\I}(q) \rho^{\mu}(t)\ .
\label{eq:supermatrix}
\end{equation}
The evolution of $\tilde{\rho}^{\mu}_{\I}(q)$ is described by 
\begin{equation}
\tilde{\rho}(q) + d\tilde{\rho}_{\I}^{\mu}(q)
=
\big[ \tilde{P}(q) + d\tilde{P}_{\I}^{\mu}(q)\big]
\big[ \rho (t) + d\rho^{\mu}(t)\big] \ .
\label{eq:rho(q)evolution}
\end{equation}
The result of this process is the joint stochastic
equation\cite{photodetection1}
%%%%%%%%%%%%%%%%%\begin{widetext}
\begin{eqnarray}
d\tilde{\rho}_{\I}^{\mu}(q)
&=& \bigg\{dt\ro{-\frac{\du}{\du q}m
+ \frac{D_\mathrm{i}}{2}\frac{\du^2}{\du q^2}}
\bigg. \nn \\
&& \bigg.
+ \ro{m - D_\mathrm{i}\frac{\du}{\du q}}\frac{\I dt}{D_\Sigma}
+ \sqrt{\frac{D_\mathrm{i}D_\mathrm{o}}{D_{\Sigma}}}
  \frac{\du}{\du q}dW'
\bigg. \nn \\
&& \bigg.
+ \mathrm{E}\sq{dN} + dt\bigg( {\cal L} - \J{{\cal T + X}\hat{n}}
\bigg)
\bigg\}\tilde{\rho}(q) \nn \\
&& +~dN\cu{\frac{\J{{\cal T + X}\hat{n}}\tilde{\rho}\ro{q-\e}}{
\mathrm{E}\sq{dN}/dt}
- \tilde{\rho}(q)} .
\label{eq:joint:qpc}
\end{eqnarray}
%%%%%%%%%%%%%%%%%\end{widetext}

Averaging over unobserved processes ($dW'$ and $dN$) is the next step
in the derivation of realistic quantum trajectory equations and yields
an expression for $d\tilde{\rho}_\I(q)$.  This procedure removes the
stochasticity associated with the unobserved processes within the
detector and leaves the stochasticity associated with the measurement
($\I$).  The resulting equation is called a superoperator Zakai
equation as we have obtained a quantum analogue of the Zakai equation
in that from measurement we are conditioning the state of a
super-system that contains a quantum system.  It is important to
realize that after averaging over unobserved processes the
super-system state $\tilde{\rho}(q) + d\tilde{\rho}_{\I}(q)$ will
\textit{not} factorize as in \erf{eq:rho(q)evolution}.

\subsection{Normalization}
\label{subsec:normalization}
Normalization of the superoperator Zakai equation is the final step in
our derivation and yields the superoperator Kushner-Stratonovich (SKS)
equation.  The normalization is achieved as follows:
\begin{equation}
\rho(q) + d\rho_{\I}(q)
= \frac{\tilde{\rho}(q) + d\tilde{\rho}_{\I}(q)}
{\int\tr{\tilde{\rho}_{\I}(q) + d\tilde{\rho}_{\I}(q)}dq}\ .
\label{eq:norm1}
\end{equation}
After normalization the true expression for the observed current $\I$
should be substituted into the SKS equation. 
The true probability distribution for $\I$ can be found using
\erf{eq:ze} in \erf{eq:P=Lambda} to yield
\begin{equation}
P(\I) = \ro{2\pi\nu}^{-1/2}\exp
\sq{-\ro{\I+\alpha\an{Q}-\beta}^2/2\nu}\ ,
\label{eq:P(I)}
\end{equation}
where $\an{Q}=\int q P(q) dq$. Thus, the true expression for the
observed current is
\begin{equation}
\label{eq:TrueI}
\I dt = \ro{-\alpha \an{Q} + \beta}dt + \sqrt{D_\Sigma}d\W,
\end{equation}
where $d\W$ is the \textit{observed} white noise (a Wiener increment).
Here the average is $\an{Q}=\int q\tr{\rho_{\I} (q)} dq$, since we are
considering the output $\I$ for the combined classical-quantum
super-system.

%%%%%%%%%%%%%%%%%%%%%%%%%%%%%
%%%%%%%%%%%%%%%%%%%%%%%%%%%%%
%%%%%%%%%%%%%%%%%%%%%%%%%%%%%
Averaging over the unobserved noise $dW'$ and tunneling process
$dN$ yields the superoperator Zakai equation, which upon normalization
via \erf{eq:norm1} and substitution of \erf{eq:TrueI} for $\I$
produces the SKS equation:
\begin{eqnarray}
d\rho_\I (q) &=& dt\sq{\frac{\du}{\du q}\ro{\alpha
    q-\beta} + \frac{D_\mathrm{i}}{2}\frac{\du^2}{\du q^2}}\rho_\I(q) \nn\\
&&-\frac{d\W}{\sqrt{D_\Sigma}}\sq{\alpha\ro{q -\an{Q}}
    +D_\mathrm{i}\frac{\du}{\du q}}\rho_\I(q)
    \nn \\ 
&&+~dt~{\cal L} \rho_\I(q) \nn\\
&&+~dt~\J{{\cal T} + {\cal X}\hat{n}}
          \sq{ \rho_\I(q-\e) - \rho_\I(q)}.
\label{eq:skse:qpc}
\end{eqnarray}
This is the main result of our paper.  The first line of
\erf{eq:skse:qpc} describes the evolution of the classical measurement
device.  The second line consists of two terms: the first term
describes information gain about the measurement device ($q$) from its
output -- \erf{eq:TrueI};  the second term describes back-action on
the classical device due to the observed noise.  The third line
describes the average evolution of the quantum system, including
quantum back-action.  The final line describes the effect of the
quantum system on the measurement device.

It is worth noting that the term involving ${\cal J}$ represents
average evolution due to electron tunneling events through the QPC.
It changes the most likely value for the charge of the parasitic
capacitor from $Q$ to $Q-\e$ when an electron tunnels through the QPC
-- effectively counting the average number of electrons passing
through the QPC.  The approach of \Ref{Shnir98} (also used in
\Ref{goan04}) involves a similar technique in which the exact number
of electrons that have tunneled through the detector is tracked.  This
was also considered in the (earlier) derivation of the rate equations
in \Ref{gurvitz}.

The numerical solution of \erf{eq:skse:qpc} would produce a trajectory
for the state of the combined circuit-qubit system conditioned by a
particular realization of the measured current $\I(t)$.  The
normalized conditioned qubit state is found from
\begin{equation}
\label{eq:returntoqubitstate}
\rho_{\I}(t) = \int \rho_{\I}(q) dq\ .
\end{equation}
Thus, the realistic quantum trajectories (for the qubit state alone)
are obtained by numerically solving the SKS equation and using
\erf{eq:returntoqubitstate}.  The results of this procedure will be
presented in a future paper.

%%%%%%%%%%%%%%%%%%%%%%%%%%%%%%%%%%%%%%%%%%%%%%%%%%%%%%%%%%%%%%%%%%%%%%%%%
\section{Discussion}
\label{sec:discussion}
A simple consistency check for our SKS equation [\erf{eq:skse:qpc}] is
to integrate it over all $q$ and recover the unconditional master
equation.  It is easy to confirm that this is indeed the case using
the fact that \textit{well-behaved} probability distributions (and
their derivatives) vanish at $\pm \infty$.

A considerably more difficult task is to attempt recovery of the ideal
conditional master equation (\ref{eq:MEqpc}) from the SKS equation
(\ref{eq:skse:qpc}).  In theory, this should be possible in the limit
of a measurement circuit with a small response time given by
$R_\mathrm{i}C$ (large bandwidth $\alpha=\ro{R_\mathrm{i}C}^{-1}$) and
low noises $D_\mathrm{i}$ and $D_\mathrm{o}$.  We now explore this
question in detail.

The time taken to determine which CQD is occupied by the qubit
electron is equal to the time required to ascertain the rate of
tunneling through the detector.  This task is made considerably more
difficult by the white noise and finite bandwidth of the circuit
containing the detector.  Without the white noise the observer would
see a spike in the current every time there was a tunneling event. 
Depending on the relative sizes of the noise, the tunneling rates and
the circuit response time, the white noise will obscure the spikes in
the current so that the observer must rely on the {\it average}
current to distinguish between the two qubit states.  The consequence
of averaging out the white noise (by integrating the current 
${\cal I}$ over some time $\tau$) is that if the qubit electron is
tunneling on a time scale $\Omega^{-1}$ shorter than $\tau$ then the
state of the quantum system cannot be followed.

We will now present an order of magnitude estimate of the effective
bandwidth of the measurement device, which is defined as the frequency
at which a signal (power) to noise (power) ratio of unity is
obtained.  Here we take the signal as being the current that flows
through the QPC when it is in the more conducting state.

To find the noise and signal power we take the Fourier transform of
Eqs. (\ref{eq:I}) and (\ref{eq:dQfinalqpc}) in order to obtain the
spectrum of the current $\I$.\cite{OxJPCM03}
  %This procedure gives results similar to those in \Ref{OxJPCM03}.  
The signal power is 
\begin{equation}
r' \e^{2}\frac{\alpha^{2}}{\alpha^{2}+\omega^{2}}
\label{eq:signalP}
\end{equation}
and the noise power is
\begin{equation}
D_\mathrm{i}\frac{\omega^{2}}{\alpha^{2}+\omega^{2}}+D_\mathrm{o}\ .
\label{eq:noiseP}
\end{equation}
Upon equating the signal and noise powers, and at this frequency
setting $\omega=\alpha_\mathrm{eff}$, we have an effective bandwidth
of
\begin{equation}
\alpha_\mathrm{eff} = \frac{\alpha}{\sqrt{\mathbf{N}}}
                   \sqrt{1-\frac{D_\mathrm{o}}{r' \e^{2}}}
\approx \frac{\alpha}{\sqrt{\mathbf{N}}}\ ,
\label{eq:effBW}
\end{equation}
where the dimensionless noise parameter $\textbf{N}$ is
defined according to
\begin{equation}
\mathbf{N} = \frac{\ro{D_\mathrm{i}+D_\mathrm{o}} }{r' \e^{2}}\ .
\end{equation}
The approximation of \erf{eq:effBW} holds in the limit where the
noise power $D_\mathrm{o}$ is small compared to the signal power
$r' \e^2$ (which is the regime that will lead to good measurements
of the qubit state).  For an observer to be able to follow the
evolution of the qubit reasonably well we must have
\mbox{$\alpha_\mathrm{eff}>\Omega$}.%
\cite{photodetection1,photodetection2}

%%%%%%%%%%%%%%%%%%%%%%%%%%%%%%%%%%%%%%%%%%%%%%%%%%%%%%%%%%%%%%%%%%%%%%
\section{Conclusion}
\label{sec:conclusion}

\subsection{Summary}
\label{subsec:summary}
We have presented a new model for continuous measurement of a coupled
quantum dot (CQD) charge qubit by a low transparency quantum point
contact (QPC).  We considered the evolution of this solid-state
qubit conditioned on the output available to a realistic observer,
which has been filtered and degraded (i.e. made more noisy) by an
external circuit.  This description is closer to the true conditioned
evolution of the system, not a hypothetical evolution conditioned on
the microscopic events occurring within the detector, to which a real
observer has no direct access.  Knowledge of the state of a quantum
system conditioned on actual measurement results is essential for
understanding and designing feedback control.\cite{WMilPRL93,
  WisePRL95, ArmenPRL02, DtyJac99, DohertyPRA00, opencontrol,
  SmithPRL02, korotkov01b, RuskovFB02}
It is also expected to be important in quantum computing, both for
state preparation and quantum error
correction.\cite{AhnWisMil03,SarAhnJacMil04,AhnWisJac04}

Our model for the conditional dynamics of the qubit due to measurement
by a low transparency quantum point contact (QPC) was based on the
quantum trajectory models of Refs.~\onlinecite{setpaper}
and~\onlinecite{qpc1}.  We have presented a stochastic master equation
that describes the time evolution of the measured qubit conditioned on
a hypothetical detector output.
Korotkov has derived equivalent conditional dynamics equations for the
QPC (with notational differences) using a Bayesian
formalism.\cite{korotkov99, korotkov01b, korotkov01c, korotkov03,
  korotkov03review}
 %For realistic measurement 
We generalized a realistic quantum trajectory
theory\cite{warwismab,photodetection1} (recently developed for 
photodetectors) to treat solid-state detectors.  The solutions of the
resulting stochastic equations are the ``realistic quantum
trajectories'' of the measured qubit state.  These will be presented
elsewhere.

%%%%%%%%%%%%%%%%%%%%%%%%%%%%%%%%%%%%%%%%%%%%%%%%%%%%%%%%%%%%%%%%%%%%%%%%%
\subsection{Comparison to previous work}
\label{subsec:previous}
The conditional dynamics of continuously monitored CQD systems has
been studied in Refs.~\onlinecite{korotkov99, korotkov01b,
  korotkov01c, setpaper, qpc1, goanmilburn, korotkov03,
  korotkov03review, gurvitzPRL}. 
However, the model of \textit{realistic} measurement that we
have presented in this paper is new.  Korotkov has recently presented
a phenomenological theory\cite{korotkov03} involving ``non-ideal''
detectors that is in the same spirit as ours (see
Appendix~\ref{sec:korotkov} for a derivation of Korotkov's result 
using our stochastic master equation approach).  However, he still
assumed an infinite bandwidth detector and also assumed that
the ideal detector could be described by diffusion rather than jumps.
We believe that our approach offers a more satisfying description of
this measurement process because the non-Markovian\footnote{The
  authors in \Ref{ruskov} take into account the effect of a finite
  bandwidth on a feedback algorithm (which has not been optimized
  because it is not based on the conditional state of the qubit). This
  is different to our work where we have calculated the conditional
  state given a bandwidth limitation in the measurement.}
effects of a realistic measurement circuit are included
and the tunneling process through the QPC is described as a point
process (jumps) as one would expect.

Finite detector temperature effects in the case of bare measurement
were not considered in our model.  The effects of a non-zero detector
temperature $T_\mathrm{d}$ have been considered
previously\cite{qpc1,korotkov01b,ruskov} and result in an
approximately linear
($\coth \sq{\e V_\mathrm{d}/2k_\mathrm{B}T_\mathrm{d}}$) increase in
the ensemble decoherence rate and shot noise level with $T_\mathrm{d}$
for $\e V_\mathrm{d} < 2k_\mathrm{B}T_\mathrm{d}$, where
$V_\mathrm{d}$ is the detector bias voltage.  With a detector
temperature of the order of mK,\cite{sct} finite temperature effects
could be expected to become important at bias voltages of
$V_\mathrm{d} < 0.3\mu V$. These voltages are several orders of
magnitude below a sample bias voltage for maximum response of a single
electron transistor (SET) detecting the charge state of a quantum
dot,\cite{MIT1997,MIT1999} which suggests that our omission of finite
detector temperature effects in the bare detector scenario is
reasonable.

%%%%%%%%%%%%%%%%%%%%%%%%%%%%%%%%%%%%%%%%%%%%%%%%%%%%%%%%%%%%%%%%%%%%
\subsection{Future Work}
\label{subsec:future}
There are many possibilities for future work in the theory of
realistic quantum trajectories.  Other detectors will be considered --
for example, the single electron transistor\cite{FultonDolan,
  AverinLikharev} (SET).  As the field of mesoscopic electronics
is progressing at such a tremendous rate it is likely that the choice
of detector will quickly become outdated.  In fact, the SET has
already been surpassed by the radio-frequency (RF)
SET\cite{electrometer,rfset2,set} as the measuring device of choice
for the read out of the charge state of a mesoscopic qubit.  This is
one reason why we view the work in this paper as preliminary.  The
extension of realistic quantum trajectories to the RF-SET is a future
aim.

Further work is also appropriate for the circuit model, which at
present is considerably simplified, but is a good starting point which
models some essential \textit{effects} of a realistic measurement.

These and other possibilities for work on the theory of realistic
quantum trajectories will be pursued in the future.

%%%%%%%%%%%%%%%%%%%%%%%%%%%%%%%%%%%%%%%%%%%%%%%%%%%%%%%%%%%%%%%%%%%%
\acknowledgments
This research was supported in part by the Australian Research Council
and the State of Queensland.  Neil Oxtoby acknowledges Kate Cor for
enlightening discussions.

%%%%%%%%%%%%%%%%%%%%%%%%%%%%%%%%%%%%%%%%%%%%%%%%%%%%%%%%%%%%%%%%%%%%
\appendix

\section{Derivation of Korotkov's result using our approach}
\label{sec:korotkov}

In this appendix we derive an equivalent result to Korotkov's
phenomenological result for non-ideal detectors\cite{korotkov03}
using a diffusive stochastic master equation approach and our
method involving observed and unobserved noises.

We begin by defining a diffusive \textit{linear} stochastic master
equation\cite{WiseQTQM} for the state matrix of a measured two level
system. The \ito stochastic master equation involving three classical,
normalized white noise processes $dW_0$ (ideal detector output noise),
$dW_1$ (extra output and back-action noise) and $dW_3$ (unobserved
back-action noise) is
\begin{eqnarray}
d\tilde{\rho} &=& dt {\cal L}\tilde{\rho}
~+~ dW_0\sqrt{\kappa_0}~\ro{\sig_z\tilde{\rho}+\tilde{\rho}\sig_z}
\nn \\
&&
+~ dW_1\sqrt{\kappa_1}~\h{-i\sig_z}\tilde{\rho}
%\nn \\
%&&
+~ dW_3\sqrt{\kappa_3}~\h{-i\sig_z}\tilde{\rho}\ ,\nn\\
\label{eq:linearSME}
\end{eqnarray}
where ${\cal L} = \h{-i\hat{H}}
+ \ro{\kappa_0 + \kappa_1 + \kappa_3}\D{\sig_z}$
and \mbox{$\hat{H} = \Omega\sig_x + \half\varepsilon\sig_z$}.
\erf{eq:Hcal} defines the superoperator ${\cal H}$.

The three white noise processes $dW_0$, $dW_1$ and $dW_3$ %were chosen to
correspond to Korotkov's three unnormalized noise processes
$\xi_0(t)$, $\xi_1(t)$ and $\xi_3(t)$. The white noise $dW_0/dt$
represents the output of an ideal detector. Korotkov's added output
noise is known as dark noise,\footnote{This terminology is from
  quantum optics. Dark noise refers to electronic noise generated
  within a detector even when no field illuminates it, i.e. in the
  dark.}
which we model by setting the output of the realistic detector to be
the current
\begin{equation}
\I (t)dt = \ro{\sqrt{\phi_0}dW_0 +
  \sqrt{\phi_1}dW_1}/\sqrt{\phi_\Sigma}\ ,
\label{eq:realI}
\end{equation}
where $\phi_0$ is the shot noise power, %(of the ideal detector output),
$\phi_1$ is the dark noise power and $\phi_\Sigma \equiv \phi_0 +
\phi_1$ ensures that $\I(t)$ has a normalized Gaussian white noise
distribution.

We now desire the quantum trajectory for the system state $\rho_\I$
conditioned on the realistic detector output in \erf{eq:realI} rather
than on $dW_0/dt$. %To proceed we rewrite $dW_0$ as
We rewrite $dW_0$ as
\begin{equation}
dW_0 = \ro{\sqrt{\phi_0}\I (t)dt +
  \sqrt{\phi_1}dW'}/\sqrt{\phi_\Sigma}\ ,
\label{eq:dW_0}
\end{equation}
where
$dW' = \ro{\sqrt{\phi_1}dW_0 - \sqrt{\phi_0}dW_1}/\sqrt{\phi_\Sigma}$
is an unobserved normalized noise process that is independent of the
observed output $\I(t)$. The extra output noise $dW_1$ can be
expressed in terms of observed and unobserved quantities as
\begin{equation}
dW_1 = \ro{\sqrt{\phi_1}\I(t)dt -
  \sqrt{\phi_0}dW'}/\sqrt{\phi_\Sigma}\ .
\label{eq:dW_1}
\end{equation}

Substituting these into \erf{eq:linearSME} yields
\begin{eqnarray}
d\tilde{\rho} &=& dt {\cal L}\tilde{\rho}
~+~ \frac{{\sqrt{\phi_0}\I (t)dt +
    \sqrt{\phi_1}dW'}}{\sqrt{\phi_\Sigma}}
\sqrt{\kappa_0}~\ro{\sig_z\tilde{\rho}+\tilde{\rho}\sig_z}
\nn \\
&&
+~\frac{{\sqrt{\phi_1}\I(t)dt -
    \sqrt{\phi_0}dW'}}{\sqrt{\phi_\Sigma}}
\sqrt{\kappa_1}~\h{-i\sig_z}\tilde{\rho}
\nn \\
&&
+~ dW_3\sqrt{\kappa_3}~\h{-i\sig_z}\tilde{\rho}\ .
\label{eq:linearSME_2}
\end{eqnarray}

Averaging over the unobserved noise processes $dW'$ and $dW_3$ removes
the terms involving $dW'$ and $dW_3$ from \erf{eq:linearSME_2}. 
%The next step in our approach is to average over the unobserved
%noises $dW'$ and $dW_3$. The result is simply \erf{eq:linearSME_2}
%with the terms involving $dW'$ and $dW_3$ absent.
%
%\begin{eqnarray}
%d\tilde{\rho} &=& dt {\cal L}\tilde{\rho}
%~+~ \I(t)dt\sqrt{\frac{\phi_0}{\phi_\Sigma}}
%\sqrt{\kappa_0}~\ro{\sig_z\tilde{\rho}+\tilde{\rho}\sig_z}
%\nn \\
%&&
%+~\I(t)dt\sqrt{\frac{\phi_1}{\phi_\Sigma}}
%\sqrt{\kappa_1}~\h{-i\sig_z}\tilde{\rho}\ .
%\label{eq:linearSME_3}
%\end{eqnarray}
%
Normalization of this result is performed in a similar manner as in
\Sec{subsec:normalization} with the final result being the
following non-linear SME:
\begin{eqnarray}
d\rho &=& dt {\cal L}\rho
+\sqrt{\frac{\kappa_0\phi_0}{\phi_\Sigma}}dt
\ro{\I(t) - 2\an{\sig_z}\sqrt{\frac{\kappa_0\phi_0}{\phi_\Sigma}}}
~\h{\sig_z}\rho
\nn \\
&&
+ dt\sqrt{\frac{\kappa_1\phi_1}{\phi_\Sigma}}
\ro{\I(t) - 2\an{\sig_z}\sqrt{\frac{\kappa_0\phi_0}{\phi_\Sigma}}}
~\h{-i\sig_z}\rho\ .\nn \\
\label{eq:nonlinearSME}
\end{eqnarray}

Now we substitute in for the actual measured current, 
$\I(t)dt = 2\an{\sig_z}\sqrt{\kappa_0\phi_0/\phi_\Sigma}dt + d\W$,
to obtain
\begin{eqnarray}
d\rho &=& dt {\cal L}\rho
+\sqrt{\frac{\kappa_0\phi_0}{\phi_\Sigma}}
~d\W
~\h{\sig_z}\rho
\nn \\
&&
+ \sqrt{\frac{\kappa_1\phi_1}{\phi_\Sigma}}
~d\W
~\h{-i\sig_z}\rho\ ,
\label{eq:SME}
\end{eqnarray}
which is equivalent to the non-ideal result of Korotkov in
\Ref{korotkov03} with some notational differences.

%Using $\hbar =1$, the notational correspondences between \Ref{korotkov03}
%and our work here, respectively, are
%\mbox{$H = \Omega$},
%\mbox{$S= \phi$}, 
%\mbox{$\xi_{0+1}(t)=\sqrt{\phi_\Sigma}d\W /dt$}, 
%\mbox{$\Delta I = -2\sqrt{\kappa_0\phi_0}$}, 
%\mbox{$A 2\sqrt{\kappa_1/\phi_1}$}, 
%\mbox{$\ro{\Delta I}^2/4S_0 = \kappa_0$}, 
%\mbox{$A^2S_1/4 = \kappa_1$}, 
%\mbox{$\gamma_3 = \kappa_3$} and
%\mbox{$\Gamma_{\Sigma} = \kappa_0+\kappa_1+\kappa_3$}.

%%%%%%%%%%%%%%%%%%%%%%%%%%%%%%%%%%%%%%%%%%%%%%%%%%%%%%%%%%%%%%%%%%%%%

\bibliography{rqtqpc}

\begin{thebibliography}{53}
\expandafter\ifx\csname natexlab\endcsname\relax\def\natexlab#1{#1}\fi
\expandafter\ifx\csname bibnamefont\endcsname\relax
  \def\bibnamefont#1{#1}\fi
\expandafter\ifx\csname bibfnamefont\endcsname\relax
  \def\bibfnamefont#1{#1}\fi
\expandafter\ifx\csname citenamefont\endcsname\relax
  \def\citenamefont#1{#1}\fi
\expandafter\ifx\csname url\endcsname\relax
  \def\url#1{\texttt{#1}}\fi
\expandafter\ifx\csname urlprefix\endcsname\relax\def\urlprefix{URL }\fi
\providecommand{\bibinfo}[2]{#2}
\providecommand{\eprint}[2][]{\url{#2}}

\bibitem[{\citenamefont{Kane}(1998)}]{kane}
\bibinfo{author}{\bibfnamefont{B.~E.} \bibnamefont{Kane}},
  \bibinfo{journal}{Nature\ (London)} \textbf{\bibinfo{volume}{393}},
  \bibinfo{pages}{133} (\bibinfo{year}{1998}).

\bibitem[{\citenamefont{Loss and DiVincenzo}(1998)}]{dots}
\bibinfo{author}{\bibfnamefont{D.}~\bibnamefont{Loss}} \bibnamefont{and}
  \bibinfo{author}{\bibfnamefont{D.~P.} \bibnamefont{DiVincenzo}},
  \bibinfo{journal}{Phys.\ Rev.\ A} \textbf{\bibinfo{volume}{57}},
  \bibinfo{pages}{120} (\bibinfo{year}{1998}).

\bibitem[{\citenamefont{Privman et~al.}(1998)\citenamefont{Privman, Vagner, and
  Kventsel}}]{privman}
\bibinfo{author}{\bibfnamefont{V.}~\bibnamefont{Privman}},
  \bibinfo{author}{\bibfnamefont{I.~D.} \bibnamefont{Vagner}},
  \bibnamefont{and} \bibinfo{author}{\bibfnamefont{G.}~\bibnamefont{Kventsel}},
  \bibinfo{journal}{Phys.\ Lett.\ A} \textbf{\bibinfo{volume}{239}},
  \bibinfo{pages}{141} (\bibinfo{year}{1998}).

\bibitem[{\citenamefont{Imamo{\= g}lu et~al.}(1999)\citenamefont{Imamo{\= g}lu,
  Awschalom, Burkard, DiVincenzo, Loss, Sherwin, and Small}}]{imamoglu}
\bibinfo{author}{\bibfnamefont{A.}~\bibnamefont{Imamo{\= g}lu}},
  \bibinfo{author}{\bibfnamefont{D.~D.} \bibnamefont{Awschalom}},
  \bibinfo{author}{\bibfnamefont{G.}~\bibnamefont{Burkard}},
  \bibinfo{author}{\bibfnamefont{D.~P.} \bibnamefont{DiVincenzo}},
  \bibinfo{author}{\bibfnamefont{D.}~\bibnamefont{Loss}},
  \bibinfo{author}{\bibfnamefont{M.}~\bibnamefont{Sherwin}}, \bibnamefont{and}
  \bibinfo{author}{\bibfnamefont{A.}~\bibnamefont{Small}},
  \bibinfo{journal}{Phys.\ Rev.\ Lett.} \textbf{\bibinfo{volume}{83}},
  \bibinfo{pages}{4204} (\bibinfo{year}{1999}).

\bibitem[{\citenamefont{Vrijen et~al.}(2000)\citenamefont{Vrijen, Yablonovitch,
  Wang, Jiang, Balandin, Roychowdhury, Mor, and DiVincenzo}}]{vrijen}
\bibinfo{author}{\bibfnamefont{R.}~\bibnamefont{Vrijen}},
  \bibinfo{author}{\bibfnamefont{E.}~\bibnamefont{Yablonovitch}},
  \bibinfo{author}{\bibfnamefont{K.}~\bibnamefont{Wang}},
  \bibinfo{author}{\bibfnamefont{H.~W.} \bibnamefont{Jiang}},
  \bibinfo{author}{\bibfnamefont{A.}~\bibnamefont{Balandin}},
  \bibinfo{author}{\bibfnamefont{V.}~\bibnamefont{Roychowdhury}},
  \bibinfo{author}{\bibfnamefont{T.}~\bibnamefont{Mor}}, \bibnamefont{and}
  \bibinfo{author}{\bibfnamefont{D.}~\bibnamefont{DiVincenzo}},
  \bibinfo{journal}{Phys.\ Rev.\ A} \textbf{\bibinfo{volume}{62}},
  \bibinfo{pages}{012306} (\bibinfo{year}{2000}).

\bibitem[{\citenamefont{DiVincenzo}(1995)}]{longcoherence}
\bibinfo{author}{\bibfnamefont{D.~P.} \bibnamefont{DiVincenzo}},
  \bibinfo{journal}{Science} \textbf{\bibinfo{volume}{270}},
  \bibinfo{pages}{255} (\bibinfo{year}{1995}).

\bibitem[{\citenamefont{Burkard and Loss}(2002)}]{burkloss2002}
\bibinfo{author}{\bibfnamefont{G.}~\bibnamefont{Burkard}} \bibnamefont{and}
  \bibinfo{author}{\bibfnamefont{D.}~\bibnamefont{Loss}},
  \bibinfo{journal}{Europhysics News} \textbf{\bibinfo{volume}{33}}
  (\bibinfo{year}{2002}).

\bibitem[{\citenamefont{Vandersypen et~al.}(2003)\citenamefont{Vandersypen,
  Hanson, {Willems van Beveren}, Elzerman, Griedanus, {De Franceschi}, and
  Kouwenhoven}}]{delftspin}
\bibinfo{author}{\bibfnamefont{L.~M.~K.} \bibnamefont{Vandersypen}},
  \bibinfo{author}{\bibfnamefont{R.}~\bibnamefont{Hanson}},
  \bibinfo{author}{\bibfnamefont{L.~H.} \bibnamefont{{Willems van Beveren}}},
  \bibinfo{author}{\bibfnamefont{J.~M.} \bibnamefont{Elzerman}},
  \bibinfo{author}{\bibfnamefont{J.~N.} \bibnamefont{Griedanus}},
  \bibinfo{author}{\bibfnamefont{S.}~\bibnamefont{{De Franceschi}}},
  \bibnamefont{and} \bibinfo{author}{\bibfnamefont{L.~P.}
  \bibnamefont{Kouwenhoven}}, in \emph{\bibinfo{booktitle}{Quantum\ Computing\
  and\ Quantum\ Bits\ in\ Mesoscopic\ Systems}}, edited by
  \bibinfo{editor}{\bibfnamefont{A.~J.} \bibnamefont{Leggett}},
  \bibinfo{editor}{\bibfnamefont{B.}~\bibnamefont{Ruggiero}}, \bibnamefont{and}
  \bibinfo{editor}{\bibfnamefont{P.}~\bibnamefont{Silvestrini}}
  (\bibinfo{publisher}{Kluwer\ Academic/Plenum\ Publishers},
  \bibinfo{address}{New\ York}, \bibinfo{year}{2003}).

\bibitem[{\citenamefont{Gurvitz}(1997)}]{gurvitz}
\bibinfo{author}{\bibfnamefont{S.~A.} \bibnamefont{Gurvitz}},
  \bibinfo{journal}{Phys.\ Rev.\ B} \textbf{\bibinfo{volume}{56}},
  \bibinfo{pages}{15215} (\bibinfo{year}{1997}).

\bibitem[{\citenamefont{Shnirman and Sch{\"o}n}(1998)}]{Shnir98}
\bibinfo{author}{\bibfnamefont{A.}~\bibnamefont{Shnirman}} \bibnamefont{and}
  \bibinfo{author}{\bibfnamefont{G.}~\bibnamefont{Sch{\"o}n}},
  \bibinfo{journal}{Phys.\ Rev.\ B} \textbf{\bibinfo{volume}{57}},
  \bibinfo{pages}{15400} (\bibinfo{year}{1998}).

\bibitem[{\citenamefont{Korotkov}(1999)}]{korotkov99}
\bibinfo{author}{\bibfnamefont{A.~N.} \bibnamefont{Korotkov}},
  \bibinfo{journal}{Phys.\ Rev.\ B} \textbf{\bibinfo{volume}{60}},
  \bibinfo{pages}{5737} (\bibinfo{year}{1999});
  \textbf{63} 085312 (2001).

\bibitem[{\citenamefont{Korotkov}(2001{\natexlab{a}})}]{korotkov01b}
\bibinfo{author}{\bibfnamefont{A.~N.} \bibnamefont{Korotkov}},
  \bibinfo{journal}{Phys.\ Rev.\ B} \textbf{\bibinfo{volume}{63}},
  \bibinfo{pages}{115403} (\bibinfo{year}{2001}{\natexlab{a}}).

\bibitem[{\citenamefont{Wiseman et~al.}(2001)\citenamefont{Wiseman, Utami, Sun,
  Milburn, Kane, Dzurak, and Clark}}]{setpaper}
\bibinfo{author}{\bibfnamefont{H.~M.} \bibnamefont{Wiseman}},
  \bibinfo{author}{\bibfnamefont{Dian Wahyu Utami}},
  \bibinfo{author}{\bibfnamefont{H.~B.} \bibnamefont{Sun}},
  \bibinfo{author}{\bibfnamefont{G.~J.} \bibnamefont{Milburn}},
  \bibinfo{author}{\bibfnamefont{B.~E.} \bibnamefont{Kane}},
  \bibinfo{author}{\bibfnamefont{A.}~\bibnamefont{Dzurak}}, \bibnamefont{and}
  \bibinfo{author}{\bibfnamefont{R.~G.} \bibnamefont{Clark}},
  \bibinfo{journal}{Phys.\ Rev.\ B} \textbf{\bibinfo{volume}{63}},
  \bibinfo{pages}{235308} (\bibinfo{year}{2001}).

\bibitem[{\citenamefont{Goan et~al.}(2001)\citenamefont{Goan, Milburn, Wiseman,
  and Sun}}]{qpc1}
\bibinfo{author}{\bibfnamefont{H.-S.} \bibnamefont{Goan}},
  \bibinfo{author}{\bibfnamefont{G.~J.} \bibnamefont{Milburn}},
  \bibinfo{author}{\bibfnamefont{H.~M.} \bibnamefont{Wiseman}},
  \bibnamefont{and} \bibinfo{author}{\bibfnamefont{H.~B.} \bibnamefont{Sun}},
  \bibinfo{journal}{Phys.\ Rev.\ B} \textbf{\bibinfo{volume}{63}},
  \bibinfo{pages}{125326} (\bibinfo{year}{2001}).

\bibitem[{\citenamefont{Korotkov}(2001{\natexlab{b}})}]{korotkov01c}
\bibinfo{author}{\bibfnamefont{A.~N.} \bibnamefont{Korotkov}},
  \bibinfo{journal}{Phys.\ Rev.\ B} \textbf{\bibinfo{volume}{64}},
  \bibinfo{pages}{193407} (\bibinfo{year}{2001}{\natexlab{b}}).

\bibitem[{\citenamefont{Goan and Milburn}(2001)}]{goanmilburn}
\bibinfo{author}{\bibfnamefont{H.-S.} \bibnamefont{Goan}} \bibnamefont{and}
  \bibinfo{author}{\bibfnamefont{G.~J.} \bibnamefont{Milburn}},
  \bibinfo{journal}{Phys.\ Rev.\ B} \textbf{\bibinfo{volume}{64}},
  \bibinfo{pages}{235307} (\bibinfo{year}{2001}).

\bibitem[{\citenamefont{Korotkov}(2003{\natexlab{a}})}]{korotkov03}
\bibinfo{author}{\bibfnamefont{A.~N.} \bibnamefont{Korotkov}},
  \bibinfo{journal}{Phys.\ Rev.\ B} \textbf{\bibinfo{volume}{67}},
  \bibinfo{pages}{235408} (\bibinfo{year}{2003}{\natexlab{a}}).

\bibitem[{\citenamefont{Korotkov}(2003{\natexlab{b}})}]{korotkov03review}
\bibinfo{author}{\bibfnamefont{A.~N.} \bibnamefont{Korotkov}}, in
  \emph{\bibinfo{booktitle}{Quantum\ Noise\ in\ Mesoscopic\ Physics}}, edited
  by \bibinfo{editor}{\bibfnamefont{Y.~V.} \bibnamefont{Nazarov}}
  (\bibinfo{publisher}{Kluwer Academic}, \bibinfo{address}{Netherlands},
  \bibinfo{year}{2003}{\natexlab{b}}), pp. \bibinfo{pages}{205--228}.

\bibitem[{\citenamefont{Gurvitz et~al.}(2003)\citenamefont{Gurvitz, Fedichkin,
  Mozyrsky, and Berman}}]{gurvitzPRL}
\bibinfo{author}{\bibfnamefont{S.~A.} \bibnamefont{Gurvitz}},
  \bibinfo{author}{\bibfnamefont{L.}~\bibnamefont{Fedichkin}},
  \bibinfo{author}{\bibfnamefont{D.}~\bibnamefont{Mozyrsky}}, \bibnamefont{and}
  \bibinfo{author}{\bibfnamefont{G.~P.} \bibnamefont{Berman}},
  \bibinfo{journal}{Phys.\ Rev.\ Lett.} \textbf{\bibinfo{volume}{91}},
  \bibinfo{pages}{066801} (\bibinfo{year}{2003}).

\bibitem[{\citenamefont{Wiseman and Milburn}(1993{\natexlab{a}})}]{WMilPRL93}
\bibinfo{author}{\bibfnamefont{H.~M.} \bibnamefont{Wiseman}} \bibnamefont{and}
  \bibinfo{author}{\bibfnamefont{G.~J.} \bibnamefont{Milburn}},
  \bibinfo{journal}{Phys.\ Rev.\ Lett.} \textbf{\bibinfo{volume}{70}},
  \bibinfo{pages}{548} (\bibinfo{year}{1993}{\natexlab{a}}).

\bibitem[{\citenamefont{Wiseman}(1995)}]{WisePRL95}
\bibinfo{author}{\bibfnamefont{H.~M.} \bibnamefont{Wiseman}},
  \bibinfo{journal}{Phys.\ Rev.\ Lett.} \textbf{\bibinfo{volume}{75}},
  \bibinfo{pages}{4587} (\bibinfo{year}{1995}).

\bibitem[{\citenamefont{Doherty and Jacobs}(1999)}]{DtyJac99}
\bibinfo{author}{\bibfnamefont{A.~C.} \bibnamefont{Doherty}} \bibnamefont{and}
  \bibinfo{author}{\bibfnamefont{K.}~\bibnamefont{Jacobs}},
  \bibinfo{journal}{Phys.\ Rev.\ A} \textbf{\bibinfo{volume}{60}},
  \bibinfo{pages}{2700} (\bibinfo{year}{1999}).

\bibitem[{\citenamefont{Doherty et~al.}(2000)\citenamefont{Doherty, Habib,
  Jacobs, Mabuchi, and Tan}}]{DohertyPRA00}
\bibinfo{author}{\bibfnamefont{A.~C.} \bibnamefont{Doherty}},
  \bibinfo{author}{\bibfnamefont{S.}~\bibnamefont{Habib}},
  \bibinfo{author}{\bibfnamefont{K.}~\bibnamefont{Jacobs}},
  \bibinfo{author}{\bibfnamefont{H.}~\bibnamefont{Mabuchi}}, \bibnamefont{and}
  \bibinfo{author}{\bibfnamefont{S.~M.} \bibnamefont{Tan}},
  \bibinfo{journal}{Phys.\ Rev.\ A} \textbf{\bibinfo{volume}{62}},
  \bibinfo{pages}{012105} (\bibinfo{year}{2000}).

\bibitem[{\citenamefont{Armen et~al.}(2002)\citenamefont{Armen, Au, Stockton,
  Doherty, and Mabuchi}}]{ArmenPRL02}
\bibinfo{author}{\bibfnamefont{M.~A.} \bibnamefont{Armen}},
  \bibinfo{author}{\bibfnamefont{J.~K.} \bibnamefont{Au}},
  \bibinfo{author}{\bibfnamefont{J.~K.} \bibnamefont{Stockton}},
  \bibinfo{author}{\bibfnamefont{A.~C.} \bibnamefont{Doherty}},
  \bibnamefont{and} \bibinfo{author}{\bibfnamefont{H.}~\bibnamefont{Mabuchi}},
  \bibinfo{journal}{Phys.\ Rev.\ Lett.} \textbf{\bibinfo{volume}{89}},
  \bibinfo{pages}{133602} (\bibinfo{year}{2002}).

\bibitem[{\citenamefont{Wiseman et~al.}(2002)\citenamefont{Wiseman, Mancini,
  and Wang}}]{opencontrol}
\bibinfo{author}{\bibfnamefont{H.~M.} \bibnamefont{Wiseman}},
  \bibinfo{author}{\bibfnamefont{S.}~\bibnamefont{Mancini}}, \bibnamefont{and}
  \bibinfo{author}{\bibfnamefont{J.}~\bibnamefont{Wang}},
  \bibinfo{journal}{Phys.\ Rev.\ A} \textbf{\bibinfo{volume}{66}},
  \bibinfo{pages}{013807} (\bibinfo{year}{2002}).

\bibitem[{\citenamefont{Smith et~al.}(2002)\citenamefont{Smith, Reiner, Orozco,
  Kuhr, and Wiseman}}]{SmithPRL02}
\bibinfo{author}{\bibfnamefont{W.~P.} \bibnamefont{Smith}},
  \bibinfo{author}{\bibfnamefont{J.~E.} \bibnamefont{Reiner}},
  \bibinfo{author}{\bibfnamefont{L.~A.} \bibnamefont{Orozco}},
  \bibinfo{author}{\bibfnamefont{S.}~\bibnamefont{Kuhr}}, \bibnamefont{and}
  \bibinfo{author}{\bibfnamefont{H.~M.} \bibnamefont{Wiseman}},
  \bibinfo{journal}{Phys.\ Rev.\ Lett.} \textbf{\bibinfo{volume}{89}},
  \bibinfo{pages}{133601} (\bibinfo{year}{2002}).

\bibitem[{\citenamefont{Ruskov and Korotkov}(2002)}]{RuskovFB02}
\bibinfo{author}{\bibfnamefont{R.}~\bibnamefont{Ruskov}} \bibnamefont{and}
  \bibinfo{author}{\bibfnamefont{A.~N.} \bibnamefont{Korotkov}},
  \bibinfo{journal}{Phys.\ Rev.\ B} \textbf{\bibinfo{volume}{66}},
  \bibinfo{pages}{041401(R)} (\bibinfo{year}{2002}).

\bibitem[{\citenamefont{Ahn et~al.}(2003)\citenamefont{Ahn, Wiseman, and
  Milburn}}]{AhnWisMil03}
\bibinfo{author}{\bibfnamefont{C.}~\bibnamefont{Ahn}},
  \bibinfo{author}{\bibfnamefont{H.~M.} \bibnamefont{Wiseman}},
  \bibnamefont{and} \bibinfo{author}{\bibfnamefont{G.~J.}
  \bibnamefont{Milburn}}, \bibinfo{journal}{Phys.\ Rev.\ A}
  \textbf{\bibinfo{volume}{67}}, \bibinfo{pages}{052310}
  (\bibinfo{year}{2003}).

\bibitem[{\citenamefont{Sarovar et~al.}(2004)\citenamefont{Sarovar, Ahn,
  Jacobs, and Milburn}}]{SarAhnJacMil04}
\bibinfo{author}{\bibfnamefont{M.}~\bibnamefont{Sarovar}},
  \bibinfo{author}{\bibfnamefont{C.}~\bibnamefont{Ahn}},
  \bibinfo{author}{\bibfnamefont{K.}~\bibnamefont{Jacobs}}, \bibnamefont{and}
  \bibinfo{author}{\bibfnamefont{G.~J.} \bibnamefont{Milburn}},
  \bibinfo{journal}{Phys.\ Rev.\ A} \textbf{\bibinfo{volume}{69}},
  \bibinfo{pages}{052324} (\bibinfo{year}{2004}).

\bibitem[{\citenamefont{Ahn et~al.}(2004)\citenamefont{Ahn, Wiseman, and
  Jacobs}}]{AhnWisJac04}
\bibinfo{author}{\bibfnamefont{C.}~\bibnamefont{Ahn}},
  \bibinfo{author}{\bibfnamefont{H.~M.} \bibnamefont{Wiseman}},
  \bibnamefont{and} \bibinfo{author}{\bibfnamefont{K.}~\bibnamefont{Jacobs}},
  \bibinfo{journal}{Phys.\ Rev.\ A} \textbf{\bibinfo{volume}{70}},
  \bibinfo{pages}{024302} (\bibinfo{year}{2004}).

\bibitem[{\citenamefont{Carmichael}(1993)}]{opensystems}
\bibinfo{author}{\bibfnamefont{H.~J.} \bibnamefont{Carmichael}},
  \emph{\bibinfo{title}{An Open Systems Approach to Quantum Optics}}
  (\bibinfo{publisher}{Springer}, \bibinfo{address}{Berlin},
  \bibinfo{year}{1993}).

\bibitem[{\citenamefont{Wiseman and Milburn}(1993{\natexlab{b}})}]{WiseMilQT93}
\bibinfo{author}{\bibfnamefont{H.~M.} \bibnamefont{Wiseman}} \bibnamefont{and}
  \bibinfo{author}{\bibfnamefont{G.~J.} \bibnamefont{Milburn}},
  \bibinfo{journal}{Phys.\ Rev.\ A} \textbf{\bibinfo{volume}{47}},
  \bibinfo{pages}{1652} (\bibinfo{year}{1993}{\natexlab{b}}).

\bibitem[{\citenamefont{Wiseman}(1996)}]{WiseQTQM}
\bibinfo{author}{\bibfnamefont{H.~M.} \bibnamefont{Wiseman}},
  \bibinfo{journal}{Quantum\ Semiclass.\ Opt.} \textbf{\bibinfo{volume}{8}},
  \bibinfo{pages}{205} (\bibinfo{year}{1996}).

\bibitem[{\citenamefont{Warszawski et~al.}(2002)\citenamefont{Warszawski,
  Wiseman, and Mabuchi}}]{warwismab}
\bibinfo{author}{\bibfnamefont{P.}~\bibnamefont{Warszawski}},
  \bibinfo{author}{\bibfnamefont{H.~M.} \bibnamefont{Wiseman}},
  \bibnamefont{and} \bibinfo{author}{\bibfnamefont{H.}~\bibnamefont{Mabuchi}},
  \bibinfo{journal}{Phys.\ Rev.\ A} \textbf{\bibinfo{volume}{65}},
  \bibinfo{pages}{023802} (\bibinfo{year}{2002}).

\bibitem[{\citenamefont{Warszawski and
  Wiseman}(2003{\natexlab{a}})}]{photodetection1}
\bibinfo{author}{\bibfnamefont{P.}~\bibnamefont{Warszawski}} \bibnamefont{and}
  \bibinfo{author}{\bibfnamefont{H.~M.} \bibnamefont{Wiseman}},
  \bibinfo{journal}{J.\ Opt.\ B:\ Quantum\ Semiclass.\ Opt.}
  \textbf{\bibinfo{volume}{5}}, \bibinfo{pages}{1}
  (\bibinfo{year}{2003}{\natexlab{a}}).

\bibitem[{\citenamefont{Field et~al.}(1993)\citenamefont{Field, Smith, Pepper,
  Ritchie, Frost, Jones, and Hasko}}]{Field}
\bibinfo{author}{\bibfnamefont{M.}~\bibnamefont{Field}},
  \bibinfo{author}{\bibfnamefont{C.~G.} \bibnamefont{Smith}},
  \bibinfo{author}{\bibfnamefont{M.}~\bibnamefont{Pepper}},
  \bibinfo{author}{\bibfnamefont{D.~A.} \bibnamefont{Ritchie}},
  \bibinfo{author}{\bibfnamefont{J.~E.~F.} \bibnamefont{Frost}},
  \bibinfo{author}{\bibfnamefont{G.~A.~C.} \bibnamefont{Jones}},
  \bibnamefont{and} \bibinfo{author}{\bibfnamefont{D.~G.} \bibnamefont{Hasko}},
  \bibinfo{journal}{Phys.\ Rev.\ Lett.} \textbf{\bibinfo{volume}{70}},
  \bibinfo{pages}{1311} (\bibinfo{year}{1993}).

\bibitem[{\citenamefont{Gardiner}(1985)}]{CWGhbook}
\bibinfo{author}{\bibfnamefont{C.~W.} \bibnamefont{Gardiner}},
  \emph{\bibinfo{title}{Handbook of Stochastic Methods for the Physical
  Sciences}} (\bibinfo{publisher}{Springer}, \bibinfo{address}{Berlin},
  \bibinfo{year}{1985}).

\bibitem[{\citenamefont{Braginsky and Khalili}(1992)}]{qmBragKhal}
\bibinfo{author}{\bibfnamefont{V.~B.} \bibnamefont{Braginsky}}
  \bibnamefont{and} \bibinfo{author}{\bibfnamefont{F.~Y.}
  \bibnamefont{Khalili}}, \emph{\bibinfo{title}{Quantum Measurement}}
  (\bibinfo{publisher}{Cambridge University Press}, \bibinfo{year}{1992}).

\bibitem[{\citenamefont{Goan}(2004)}]{goan04}
\bibinfo{author}{\bibfnamefont{H.-S.} \bibnamefont{Goan}},
  \bibinfo{journal}{Phys.\ Rev.\ B} \textbf{\bibinfo{volume}{70}},
  \bibinfo{pages}{075305} (\bibinfo{year}{2004}).

\bibitem[{\citenamefont{Schoelkopf et~al.}(2003)\citenamefont{Schoelkopf,
  Clerk, Girvin, Lehnert, and Devoret}}]{yalespie}
\bibinfo{author}{\bibfnamefont{R.~J.} \bibnamefont{Schoelkopf}},
  \bibinfo{author}{\bibfnamefont{A.~A.} \bibnamefont{Clerk}},
  \bibinfo{author}{\bibfnamefont{S.~M.} \bibnamefont{Girvin}},
  \bibinfo{author}{\bibfnamefont{K.~W.} \bibnamefont{Lehnert}},
  \bibnamefont{and} \bibinfo{author}{\bibfnamefont{M.~H.}
  \bibnamefont{Devoret}}, in \emph{\bibinfo{booktitle}{Noise and Information in
  Nanoelectronics, Sensors and Standards}}, edited by
  \bibinfo{editor}{\bibfnamefont{L.~B.} \bibnamefont{Kish}},
  \bibinfo{editor}{\bibfnamefont{F.}~\bibnamefont{Green}},
  \bibinfo{editor}{\bibfnamefont{G.}~\bibnamefont{Iannaccone}},
  \bibnamefont{and} \bibinfo{editor}{\bibfnamefont{J.~R.} \bibnamefont{Vig}}
  (\bibinfo{organization}{SPIE}, \bibinfo{year}{2003}), vol.
  \bibinfo{volume}{5115}, pp. \bibinfo{pages}{356--376}.

\bibitem[{\citenamefont{Milburn and Sun}(1998)}]{HeBiGerard}
\bibinfo{author}{\bibfnamefont{G.~J.} \bibnamefont{Milburn}} \bibnamefont{and}
  \bibinfo{author}{\bibfnamefont{H.~B.} \bibnamefont{Sun}},
  \bibinfo{journal}{Contemporary Physics} \textbf{\bibinfo{volume}{39, 1}},
  \bibinfo{pages}{67} (\bibinfo{year}{1998}).

\bibitem[{\citenamefont{Oxtoby et~al.}(2003)\citenamefont{Oxtoby, Sun, and
  Wiseman}}]{OxJPCM03}
\bibinfo{author}{\bibfnamefont{N.~P.} \bibnamefont{Oxtoby}},
  \bibinfo{author}{\bibfnamefont{H.-B.} \bibnamefont{Sun}}, \bibnamefont{and}
  \bibinfo{author}{\bibfnamefont{H.~M.} \bibnamefont{Wiseman}},
  \bibinfo{journal}{J.\ Phys.:\ Condens.\ Matter}
  \textbf{\bibinfo{volume}{15}}, \bibinfo{pages}{8055} (\bibinfo{year}{2003}).

\bibitem[{\citenamefont{Box and Tiao}(1973)}]{bayes}
\bibinfo{author}{\bibfnamefont{G.~E.~P.} \bibnamefont{Box}} \bibnamefont{and}
  \bibinfo{author}{\bibfnamefont{G.~C.} \bibnamefont{Tiao}},
  \emph{\bibinfo{title}{Bayesian Inference in Statistical Analysis}}
  (\bibinfo{publisher}{Addison-Wesley}, \bibinfo{address}{Sydney},
  \bibinfo{year}{1973}).

\bibitem[{\citenamefont{Warszawski and
  Wiseman}(2003{\natexlab{b}})}]{photodetection2}
\bibinfo{author}{\bibfnamefont{P.}~\bibnamefont{Warszawski}} \bibnamefont{and}
  \bibinfo{author}{\bibfnamefont{H.~M.} \bibnamefont{Wiseman}},
  \bibinfo{journal}{J.\ Opt.\ B:\ Quantum\ Semiclass.\ Opt.}
  \textbf{\bibinfo{volume}{5}}, \bibinfo{pages}{15}
  (\bibinfo{year}{2003}{\natexlab{b}}).

\bibitem[{\citenamefont{Ruskov and Korotkov}(2003)}]{ruskov}
\bibinfo{author}{\bibfnamefont{R.}~\bibnamefont{Ruskov}} \bibnamefont{and}
  \bibinfo{author}{\bibfnamefont{A.~N.} \bibnamefont{Korotkov}},
  \bibinfo{journal}{Phys.\ Rev.\ B} \textbf{\bibinfo{volume}{67}},
  \bibinfo{pages}{075303} (\bibinfo{year}{2003}).

\bibitem[{\citenamefont{Grabert and Devoret}(1992)}]{sct}
\bibinfo{editor}{\bibfnamefont{H.}~\bibnamefont{Grabert}} \bibnamefont{and}
  \bibinfo{editor}{\bibfnamefont{M.~H.} \bibnamefont{Devoret}}, eds.,
  \emph{\bibinfo{title}{Single Charge Tunneling: Coulomb Blockade Phenomena in
  Nanostructures}} (\bibinfo{publisher}{Plenum Press}, \bibinfo{address}{New
  York}, \bibinfo{year}{1992}).

\bibitem[{\citenamefont{Berman et~al.}(1997)\citenamefont{Berman, Zhitenev,
  Ashoori, Smith, and Melloch}}]{MIT1997}
\bibinfo{author}{\bibfnamefont{D.}~\bibnamefont{Berman}},
  \bibinfo{author}{\bibfnamefont{N.~B.} \bibnamefont{Zhitenev}},
  \bibinfo{author}{\bibfnamefont{R.~C.} \bibnamefont{Ashoori}},
  \bibinfo{author}{\bibfnamefont{H.~I.} \bibnamefont{Smith}}, \bibnamefont{and}
  \bibinfo{author}{\bibfnamefont{M.~R.} \bibnamefont{Melloch}},
  \bibinfo{journal}{J.\ Vac.\ Sci.\ Technol.\ B} \textbf{\bibinfo{volume}{15}},
  \bibinfo{pages}{2844} (\bibinfo{year}{1997}).

\bibitem[{\citenamefont{Berman et~al.}(1999)\citenamefont{Berman, Zhitenev,
  Ashoori, and Shayegan}}]{MIT1999}
\bibinfo{author}{\bibfnamefont{D.}~\bibnamefont{Berman}},
  \bibinfo{author}{\bibfnamefont{N.~B.} \bibnamefont{Zhitenev}},
  \bibinfo{author}{\bibfnamefont{R.~C.} \bibnamefont{Ashoori}},
  \bibnamefont{and} \bibinfo{author}{\bibfnamefont{M.}~\bibnamefont{Shayegan}},
  \bibinfo{journal}{Phys.\ Rev.\ Lett.} \textbf{\bibinfo{volume}{82}},
  \bibinfo{pages}{161} (\bibinfo{year}{1999}).

\bibitem[{\citenamefont{Fulton and Dolan}(1987)}]{FultonDolan}
\bibinfo{author}{\bibfnamefont{T.~A.} \bibnamefont{Fulton}} \bibnamefont{and}
  \bibinfo{author}{\bibfnamefont{G.~J.} \bibnamefont{Dolan}},
  \bibinfo{journal}{Phys.\ Rev.\ Lett.} \textbf{\bibinfo{volume}{59}},
  \bibinfo{pages}{109} (\bibinfo{year}{1987}).

\bibitem[{\citenamefont{Averin and Likharev}(1986)}]{AverinLikharev}
\bibinfo{author}{\bibfnamefont{D.~V.} \bibnamefont{Averin}} \bibnamefont{and}
  \bibinfo{author}{\bibfnamefont{K.~K.} \bibnamefont{Likharev}},
  \bibinfo{journal}{J.\ Low\ Temp.\ Phys.} \textbf{\bibinfo{volume}{62}},
  \bibinfo{pages}{345} (\bibinfo{year}{1986}).

\bibitem[{\citenamefont{Schoelkopf et~al.}(1998)\citenamefont{Schoelkopf,
  Wahlgren, Kozhevnikov, Delsing, and Prober}}]{electrometer}
\bibinfo{author}{\bibfnamefont{R.~J.} \bibnamefont{Schoelkopf}},
  \bibinfo{author}{\bibfnamefont{P.}~\bibnamefont{Wahlgren}},
  \bibinfo{author}{\bibfnamefont{A.~A.} \bibnamefont{Kozhevnikov}},
  \bibinfo{author}{\bibfnamefont{P.}~\bibnamefont{Delsing}}, \bibnamefont{and}
  \bibinfo{author}{\bibfnamefont{D.~E.} \bibnamefont{Prober}},
  \bibinfo{journal}{Science} \textbf{\bibinfo{volume}{280}},
  \bibinfo{pages}{1238} (\bibinfo{year}{1998}).

\bibitem[{\citenamefont{Fujisawa and Hirayama}(2000)}]{rfset2}
\bibinfo{author}{\bibfnamefont{T.}~\bibnamefont{Fujisawa}} \bibnamefont{and}
  \bibinfo{author}{\bibfnamefont{Y.}~\bibnamefont{Hirayama}},
  \bibinfo{journal}{Appl.\ Phys.\ Lett.} \textbf{\bibinfo{volume}{77}},
  \bibinfo{pages}{543} (\bibinfo{year}{2000}).

\bibitem[{\citenamefont{Devoret and Schoelkopf}(2000)}]{set}
\bibinfo{author}{\bibfnamefont{M.~H.} \bibnamefont{Devoret}} \bibnamefont{and}
  \bibinfo{author}{\bibfnamefont{R.~J.} \bibnamefont{Schoelkopf}},
  \bibinfo{journal}{Nature} \textbf{\bibinfo{volume}{406}},
  \bibinfo{pages}{1039} (\bibinfo{year}{2000}).

\end{thebibliography}

\end{document}